\documentclass{aastex6}
\usepackage[utf8]{inputenc}

\shorttitle{Prospects for CTA observations of RX~J1713.7$-$3946}
\shortauthors{Acero et al.}
 
\begin{document}


\title{
  Prospects for Cherenkov Telescope Array Observations of the Young Supernova Remnant RX~J1713.7$-$3946
}

\author{
F.~Acero\altaffilmark{1},
R.~Aloisio\altaffilmark{2,3},
J.~Amans\altaffilmark{4},
E.~Amato\altaffilmark{2},
L.A.~Antonelli\altaffilmark{5},
C.~Aramo\altaffilmark{6},
T.~Armstrong\altaffilmark{7},
F.~Arqueros\altaffilmark{8},
K.~Asano\altaffilmark{9},
M.~Ashley\altaffilmark{10}
% ...
}

\altaffiltext{1}{CEA/IRFU/SAp, CEA Saclay, Bat 709, Orme des Merisiers, 91191 Gif-sur-Yvette, France}
\altaffiltext{2}{Osservatorio Astrofisico di Arcetri, Largo E. Fermi, 5 - 50125 Firenze, Italy}
\altaffiltext{3}{INFN Dipartimento di Scienze Fisiche e Chimiche - Università degli Studi dell'Aquila and Gran Sasso Science Institute, Via Vetoio 1, Viale Crispi 7, 67100 L'Aquila, Italy}
\altaffiltext{4}{LUTH and GEPI, Observatoire de Paris, CNRS, PSL Research University, 5 place Jules Janssen, 92190, Meudon, France}
\altaffiltext{5}{INAF - Osservatorio Astronomico di Roma, Via di Frascati 33, 00040, Monteporzio Catone, Italy}
\altaffiltext{6}{INFN Sezione di Napoli, Via Cintia, ed. G - 80126 Napoli, Italy}
\altaffiltext{7}{Dept. of Physics and Centre for Advanced Instrumentation, Durham University, South Road, Durham DH1 3LE, United Kingdom}
\altaffiltext{8}{Grupo de Altas Energías, Universidad Complutense de Madrid., Av Complutense s/n, 28040 Madrid, Spain}
\altaffiltext{9}{Institute for Cosmic Ray Research, University of Tokyo, 5-1-5, Kashi-wanoha, Kashiwa, Chiba 277-8582, Japan}
\altaffiltext{10}{School of Physics, University of New South Wales, Sydney NSW 2052, Australia}
% ...


\def\d{{\rm d}}
\def\p{\partial}
\def\w{\wedge}
\def\o{\otimes}
\def\f{\frac}
\def\tr{{\rm tr}}
\def\Half{\frac{1}{2}}
\def\half{{\scriptstyle \frac{1}{2}}}
\def\T{\tilde}
\def\RA{\rightarrow}
\def\N{\nonumber}
\def\n{\nabla}
\def\bb{\bibitem}
\def\BE{\begin{equation}}
\def\EE{\end{equation}}
\def\BEA{\begin{eqnarray}}
\def\EEA{\end{eqnarray}}
\def\L{\label}
\def\zero{{\scriptscriptstyle 0}}

\begin{abstract}
We perform simulations for future Cherenkov Telescope Array (CTA) observations of RX~J1713.7$-$3946, a young supernova remnant (SNR) 
and one of the brightest sources ever discovered in very-high-energy (VHE) gamma rays. 
Special attention is paid to explore possible spatial (anti-)correlations of 
gamma rays with emission at other wavelengths, in particular X-rays and CO/H{\sc i} emission. 
We present a series of simulated images of
RX J1713.7$-$3946 for CTA based on 
a set of observationally motivated models for the gamma-ray emission. 
In these models, VHE gamma rays produced by high-energy electrons
are assumed to trace the non-thermal X-ray emission observed by {\it XMM-Newton},
whereas those originating from relativistic protons delineate the local gas distributions.
The local atomic and molecular gas distributions are deduced by the NANTEN team
from CO and H{\sc i} observations.  
Our primary goal is to show how one can distinguish the emission mechanism(s) of the gamma rays
(i.e., hadronic vs leptonic, or a mixture of the two) through information provided 
by their spatial distribution, spectra, and time variation. 
This work is the first attempt to quantitatively evaluate the capabilities of CTA to achieve various proposed scientific goals by observing this important cosmic particle accelerator.
\end{abstract}




\clearpage

\section{Introduction}

 \subsection{Origin of Galactic cosmic rays}

The origin of Galactic cosmic rays (CRs) protons with energies up to $10^{15.5}$eV (the so-called ``knee''), 
has been one of the long-standing problems in astrophysics \citep[e.g.,][]{drury12,blasi13,amato14}.
At present, young SNRs are the most probable candidates for being the major accelerators of CRs, 
which are sometimes also thought to have been potential ``PeVatrons''
(i.e., accelerators capable of producing charged particles at PeV scale).
CRs of heavier species like iron may reach energies near the ``second knee'' at around $10^{17}$~eV. 
The detections of synchrotron X-rays in some SNRs have already shown evidence 
for the acceleration of electrons to ultra-relativistic energies at SNR shocks 
\citep{koyama95}.
On the other hand, 
there remains the unresolved issue as to how efficiently SNRs are 
accelerating high-energy protons compared to electrons.
So far, VHE gamma-ray observations have revealed the existence of high-energy particles 
at the shock of young SNRs \citep[e.g.,][see also Ferrand \& Safi-Harb 2012]{enomoto02,aharonian04,aharonian05c,
katagiri05,hess06,hess07,magic07CasA,hess07VelaJr,hess09rcw86,hess10sn1006,hess11rxj1731,veritas11tycho}.
Although the gamma rays are suggested to originate from either leptonic (low-energy photons up-scattered by high-energy electrons) 
or hadronic ($\pi^0$-decay photons generated by accelerated protons colliding with surrounding gas) processes, 
it is generally a non-trivial task to distinguish these processes despite abundant multi-wavelength studies.
As a result, we cannot yet provide an unequivocal proof for the paradigm that Galactic CRs are predominantly produced by young SNRs. 
It has been shown that the gamma-ray spectra of the middle-aged SNRs IC~443 and W44 
have a sharp cutoff at low energies ($\sim100$~MeV), 
reminiscent of the $\pi^0$ bump that provides a direct proof for a hadronic origin of the gamma rays \citep{giuliani11,ackermann13}. 
However, the maximum CR energies inferred from the gamma rays detected in these SNRs are 
much lower than the ``knee'', 
and their gamma-ray fluxes at VHE energies are very low \citep[see e.g.,][for detailed emission models]{uchiyama10, lee2015}.
We hence expect that the younger SNR population is more qualified to be PeVatron candidates.

The acceleration mechanisms of CRs have also been studied for a long time. 
As of today, the most plausible physical process is 
suggested to be the diffusive shock acceleration (DSA) mechanism \citep{blandford87}.
This is supported by the observational fact that some young SNRs are 
found to be gamma-ray bright around their shock fronts.
Recent development of the DSA theory has revealed that 
the back-reaction of the accelerated CRs, that is their pressure against the incoming super-Alfv\'{e}n gas flow in the shock frame, 
cannot be neglected
if a large number of nuclear particles are accelerated to relativistic energies \citep[e.g.,][]{drury83,berezhko99,malkov01}. 
This can lead to strong modifications of the shock structure and a nonlinear coupling between the shock flow and CR acceleration.
Some observational results are consistent with the predictions of such nonlinear DSA (NLDSA) models 
\citep{vink03, bamba03b, bamba05a, bamba05b, warren2005, uchiyama07, helder2009}.
However, whether these models can fully explain all aspects 
revealed by the accumulating multi-wavelength observations still remains in doubt.
For example, NLDSA models often predict a ``concave'' curvature in the CR spectrum
in which the photon index decreases as CR energy increases.
In such a case, the spectral index, $s$, of the accelerated particles around the shock can be harder than 2.0 
near the maximum energy if the acceleration is highly efficient
\footnote{
We consider a CR energy spectrum in the form of ${\rm d}N(E)/{\rm d}E \propto E^{-s}$, where $E$ is the CR energy.
} 
\citep{malkov97, berezhko99, kang01}.
This prediction appears to contradict the CR spectral indices of 
$s \approx {2.2-2.4}$ inferred from recent gamma-ray observations,
radio spectral indices, and the CR spectrum at Earth. 
Recent modifications of the NLDSA theory did manage to reproduce indices softer 
than $s=2.0$ by, for example, invoking feedback effects from the 
self-generated magnetic turbulence in the shock precursor \citep[e.g.,][]{caprioli09, ferrand14b}.
Shock obliqueness and momentum dependence of the diffusion coefficient might be important for producing a softer index \citep[e.g.,][]{ferrand14}.
Another possible scenario is to consider the presence of neutral hydrogen in the
acceleration region \citep{ohira09, ohira12, blasi12, morlino13}
But the story is obviously far from complete before we reach 
an entire understanding of the plasma physics around strong collisionless shocks. 

Furthermore, the important process of CR escape into the interstellar medium (ISM) is also uncertain.
In general, the gamma-ray spectrum of young SNRs starts to decline around 10~TeV,
so that the maximum energy of CRs 
is around 30--100~TeV.
This is approximately two orders-of-magnitude lower than the knee energy.
Therefore, to explain the knee feature in the CR spectrum at Earth,
very young SNRs of ages $\lesssim100$~yr are 
anticipated to experience a ``PeVatron phase'' \citep{gabici07}
during which the highest-energy CRs are
generated under strong magnetic fields associated with high-velocity shocks \citep{voelk05} and released into the ISM.
However, we do not yet understand in detail 
when and how these high-energy particles escape from
their acceleration sites to become Galactic CRs.


\subsection{Origin of gamma rays from SNR RX~J1713.7$-$3946}

RX~J1713.7$-$3946 (also known as a radio SNR G347.3$-$0.5 \citep{slane99}), one of the 
brightest VHE gamma-ray sources ever detected 
\citep{muraishi00,enomoto02,aharonian04,hess06,hess07},
is an ideal target to study these unresolved mysteries.
The distance and age of RX~J1713.7$-$3946 are estimated to be 0.9--1.3~kpc and $\sim 1600$~yrs, respectively \citep{fukui03,moriguchi05},
which is consistent with its connection with the guest star AD393 \citep[e.g.,][]{wang97}.
This age estimate is supported by its fast shock velocity \citep{katsuda15} and the similarities of its other observed properties to other young SNRs.
%
So far, in comparison with other young shell-type SNRs, the VHE 
gamma-ray spectrum of RX~J1713.7$-$3946 is the most precisely measured 
over a wide energy band (from 0.3 to 100 TeV) thanks to its high brightness. 
In addition to VHE gamma rays,
observations in other wavebands are also available. These include
lower-energy gamma rays detected by the \textit{Fermi} Large Area Telescope ({\it Fermi}/LAT),
synchrotron radio emission and X-rays, and radio line emission from CO molecules and H{\sc i} gas.
%
Although weak thermal X-ray emission has recently been detected from the SNR interior \citep{katsuda15},
the X-ray emission is still dominated by synchrotron radiation,
which links directly to the existence of high-energy electrons.
Radio observations of CO and H{\sc i} gas have revealed 
a highly inhomogeneous medium surrounding the SNR, 
such as clumpy molecular clouds
\citep{fukui03,fukui12,fukui13,moriguchi05,sano10,sano13,sano14}.
Another radio observation of CS also confirmed the existence of very dense  ($>10^5$~cm$^{-3}$) ISM core towards the SNR \citep{maxted12}.
We are also aware that some of these characteristics are common
among several other young SNRs, including
RX~J0852.0$-$4622 \citep{katagiri05,aharonian05c,hess07VelaJr}, 
RCW~86 \citep{hess09rcw86}, and 
HESS~J1731$-$347 \citep{hess11rxj1731}.
It is noteworthy that no thermal X-ray emission has been firmly detected in these SNRs \citep[e.g.,][]{koyama97,tsunemi00,bamba12}.

Prior to the {\it Fermi} era,
the VHE gamma-ray spectrum of RX~J1713.7$-$3946 above 300~GeV
measured by H.E.S.S. was often suggested to be best explained by
a hadronic model \citep{hess06,hess07} in which the 
gamma rays originate from the decay of $\pi ^0$ mesons 
produced by CR protons interacting with the local gas. 
However, as {\it Fermi}/LAT measured the gamma-ray spectrum of RX~J1713.7$-$3946
in the 3--300~GeV energy range, it turned out that the photon index ($1.5\pm0.1$) is one more typically expected 
from leptonic (i.e., inverse-Compton) emission
from high-energy electrons with a spectral index of $s=2.0$ \citep{abdo11rxj1713}.
{\it Fermi} also reported a smooth connection between the hard GeV spectrum and the TeV domain
with a spectral index of $2.02\pm0.23$ for $E>50$~GeV up to 1~TeV \citep{ackermann16}.
The conclusion on the leptonic origin of the gamma rays from this remnant has been reached by 
recent theoretical modelings \citep[e.g.,][]{ellison10,lee2012}.
In this case, the flux ratio between synchrotron X-rays and inverse-Compton
gamma rays suggests a magnetic field strength of $B\sim10$--20~$\mu$G.
At first sight, this kind of value is inconsistent with the simple interpretation for
the observed time variability \citep{uchiyama07} and thin filamentary structures 
\citep{bamba05a,parizot06} of synchrotron X-rays through a fast energy loss of high-energy electrons,
which typically requires $B\gtrsim100$~$\mu$G.
Recent X-ray observations of the SNR RX~J0852.0$-$4622,
which show very similar features as RX~J1713.7$-$3946, 
have however recovered a shallow steepening of the synchrotron index 
behind the shock as a function of radius, 
suggesting an average magnetic field near the shock of only a few~$\mu$G \citep{Kishishita13}. 
This has been shown to be consistent with theoretical models \citep{lee2013}. 
It has also been proposed 
that a fast X-ray time variability does not necessarily require high magnetic fields 
in extended regions behind the shock, 
For example \citet{bykov08} has shown that 
a steeply falling electron distribution in a turbulent magnetic field 
can produce intermittent synchrotron emission consistent with the observations.
The thin filament of synchrotron X-rays might be explained if the downstream
magnetohydrodynamic turbulence damps exponentially \citep{pohl05}, 
although a critical assessment has been
given for the case of RX~J1713.7$-$3946 \citep{marcowith2010}.
It is worth noticing that in the framework of one-zone leptonic model the IC emission on CMB photons only cannot explain the observed gamma-ray flux which, on the contrary, requires a high level of IR background radiation about 20 times larger than the Galactic average at the position of RXJ 1713.7$-$3946 (see \citet{morlino09}
These seemingly contradictory observational facts and competing interpretations have led to much uncertainty in the quest to unravel the true origin of the gamma-ray emission from non-thermal SNRs like RX~J1713.7$-$3946. 
This age-old issue awaits a final resolution by the utilization of observatories with higher performance in the near future.   

Alternatively,
there are possibilities that a hadronic model for the gamma-ray emission of RX~J1713.7$-$3946 remains viable despite its hard spectrum.
For example, in extreme cases, NLDSA models can produce
a proton index of 1.5 \citep{yamazaki09}, resulting in a hard slope of the hadronic gamma rays 
consistent with the {\it Fermi} observation.
Another plausible scenario can be pictured by considering the effect of
shock-cloud interaction which is strongly suggested by recent 
results of CO and H{\sc i} observations towards RX~J1713.7$-$3946 \citep{fukui12}.
The higher is the energy of the CR protons,
the deeper they can penetrate into the central (i.e., high-density) part of the dense cloud cores inside a highly inhomogeneous gas environment.
Depending on the total mass fraction and the energy dependence of the CR penetration depth 
in such dense cores in a clumpy medium interacting with the shock,
an overall hard gamma-ray spectrum consistent 
with the {\it Fermi}/LAT and H.E.S.S. data can be realized \citep{zirakashvili10, inoue12, gabici14}.
On the other hand, it has been argued that since a hadronic model requires a high ambient gas density 
in order for the gamma-ray emission to be dominated by the $\pi^0$-decay component, 
bright thermal X-ray emission should be expected from the shocked gas 
which is nevertheless not detected so far \citep[e.g.,][]{ellison12}. 
In a shock-cloud interaction scenario, however, 
it is quite possible that the dense clumps of gas swept up by the blastwave can remain ``intact'' and stay at a low average ionization state and temperature.
If these dense clumps bear a significant mass fraction behind the shock, faint thermal X-rays can be naturally explained \citep{inoue12}. 

Furthermore, a strong magnetic field ($B\sim0.1$--1~mG)
can be automatically generated by the turbulent dynamo downstream of the shock that is interacting with a clumpy medium, even without considering the amplification of turbulent magnetic field driven by the streaming CR protons in the shock precursor.
A potential difficulty one can face is that, 
if $B\sim{\rm mG}$ and the gamma-ray emission is mostly hadronic, 
to explain the measured flux of radio synchrotron emitted by
electrons, the primary electron-to-proton ratio at the SNR should be
anomalously small, i.e., $K_{ep} \sim 10^{-6}$ \citep{uchiyama03,butt08}.
This is far below the observed value at Earth and the typical estimated values in
the nearby galaxies \citep{katz08}.
This might be resolved if the electrons are accelerated in the
later stages of SNR evolution relative to the protons so that
$K_{ep}$ can be different from the present value \citep{tanaka08},
although this is still highly speculative and further discussions are necessary.

Another possibility that may offer an explanation to the gamma-ray spectrum 
is that the gamma rays are contributed by a roughly comparable
mixture of leptonic and hadronic components. However, this scenario
seems unlikely because of the apparently energy-independent gamma-ray morphology
\citep{hess06}. The smooth, single-peak spectral shape without any features
would require fine-tuning of model parameters describing the two populations.
We should also note that the accurate determination of the mix of leptonic and hadronic  components is made difficult by the H.E.S.S. point spread function (PSF).

Our best hope to make progress on our understanding of this very important 
CR accelerator relies upon better data from powerful future telescopes.
The Cherenkov Telescope Array (CTA) is a next-generation observatory of 
imaging air Cherenkov telescopes (IACTs),
which consists of an array of large, medium, and small-sized IACTs 
distributed over a km$^2$ area \citep{actis2011cta,ctaconcept}.
The CTA will achieve better angular resolution and higher sensitivity in a wider energy range than the current IACT generation (see the following section). 
It is expected that we can obtain the most stringent constraints 
on theoretical models through their comparison with CTA data on RX~J1713.7$-$3946
and multi-wavelength observations of this intriguing object.

In this paper, we study prospects for CTA observations of RX~J1713.7$-$3946.
In section~2, we summarize the scientific objectives related to the capabilities of CTA .
In section~3, we describe in depth our simulation models and the results.
Discussions and summary can be found in section~4.


\section{Linking CTA capabilities with key scientific goals}

The identification of the emission mechanism of gamma rays from
non-thermal SNRs like RX~J1713.7$-$3946 is of utmost importance for advancing cosmic-ray physics.
We will evaluate the capabilities of CTA on serving this purpose as follows.
First we perform morphological studies
in order to assess the possibility of using spatial information to pin down the major component of the VHE gamma-ray emission from leptonic and hadronic origins.
In the case where the leptonic component is dominant, we can proceed to check the capability of CTA to search for a possible `additional' hadronic component using spectral analyses.
We will also evaluate the prospect of detecting a time variation of the spectral cutoff energy
over a relatively long time scale, which we believe is a very useful and unique approach for identifying the VHE emission mechanism.

\begin{itemize}
\item{\it Imaging with good spatial resolution}\\
CTA is designed to reach angular resolutions of arcminute scale at TeV energies,
which is at least a factor of three better than H.E.S.S. \citep{actis2011cta}. 
This improvement is critical in our present study
because the resolution of CTA will catch up with that of current CO and H{\sc i} gas 
observations in the radio waveband and also get closer to that of X-ray images such as those obtained by {\it XMM-Newton}.
%
Synchrotron X-rays are enhanced around the CO and H{\sc i} clumps on a length scale of 
a few~pc,\footnote{
$1~{\rm pc} \approx 0.06$~deg at a distance of 1~kpc.
}
but are relatively faint inside the clumps on a sub-pc scale. 
In other words, the peaks in X-ray brightness 
are offset from those of the CO/H{\sc i} gas distribution \citep{sano13}.
Indeed, there exist many gas clumps whose associated X-ray peaks are located
more than 0.05~deg away from their centers.
If the gamma rays are leptonic in origin and if the energy spectrum of the CR protons are spatially uniform,
we expect that the gamma rays possess brightness peaks coincident with the CO/H{\sc i} gas clumps and 
show similar spatial offsets to the X-rays.

\item{\it Broadband gamma-ray spectrum}\\
CTA will cover photon energies from 20~GeV to 100~TeV with a sensitivity about 10 times better
than H.E.S.S. \citep{actis2011cta}. 
The spectrum of RX~J1713.7$-$3946 measured by H.E.S.S. is well described by a single
smooth component. However, it is possible that 
there exists another dimmer, and harder, component at the
high-energy end of the spectrum that H.E.S.S.  
cannot discern due to its insufficient effective area.
More specifically, if the gamma-ray spectrum is described by an
inverse-Compton component with a primary electron index $s=2.0$, which has been
inferred from the {\it Fermi}/LAT observation \citep{abdo11rxj1713}, 
there can exist an accompanying hadronic component
from the CR protons with the same spectral index.
This hadronic component should appear flat in $\nu F_\nu$ space, and
its expected flux is
\begin{equation}
\nu F_\nu \approx 3\times10^{-12}~{\rm erg}~{\rm cm}^{-2}{\rm s}^{-1}
\left(
\frac{E_{\rm CR}}{3\times10^{49}~{\rm erg}}
\right)
\left(
\frac{n}{1~{\rm cm}^{-3}}
\right)~~,
\label{eqn:hadron_flux}
\end{equation} 
where we have assumed the distance to RX~J1713.7$-$3946 to be 1~kpc.
The quantities $E_{\rm CR}$ and $n$ are the total CR proton energy in the emission sites and 
the average target gas density, respectively.
By adopting the values $E_{\rm CR}=3\times 10^{49}$~ergs and $n=1$~cm$^{-3}$ as in Equation~(\ref{eqn:hadron_flux}), 
this flux is about 10~\% of the presently observed $\nu F_\nu$ spectral peak
at a photon energy of $\sim{\rm TeV}$; such a hard component can dominate at the high-energy end
of the gamma-ray spectrum and will be discernible as a spectral hardening feature by CTA.

Note that a dim hadronic component can be dominant at the highest 
gamma-ray energies only if the maximum energy $E_{\rm max}$ of the parent CR proton
spectrum is not lower than $\sim100$~TeV, and that such a high value of
$E_{\rm max}$ is not guaranteed.
In particular, if we assume the simplest one-zone model with the 
diffusive shock acceleration around the shock of RX~J1713.7$-$3946 as well as
an inverse Compton origin for the gamma rays, then 
the maximum electron energy, $\sim70$~TeV, is not limited by 
energy loss but by the SNR age or CR escape to far upstream
because of the weak average magnetic field strength of $\sim10$~$\mu$G required for a leptonic scenario,
so that the maximum proton energy is basically similar to
that of electrons.
However, the emergence of a dim hadronic emission component 
with a maximum energy of more than 50~TeV is still possible
if we consider a multi-zone model.
Multi-zone models may be motivated by the observational results referred in section 1.2, 
which seem to be inconsistent with each other
\footnote{For instance, if the gamma-ray emission is leptonic, 
the magnetic field is weaker than those suggested by rapid variability and thin filaments of synchrotron X-rays (if their origins are associated to fast electron cooling near the cutoff which may or may not be the case).
Another example is that the number density of the local gas should be small to explain
the non-detection of thermal X-ray lines. This situation may be unlikely if we take into account the
fact that this SNR is apparently interacting with dense molecular clouds, although it
depends much on the timing of the shock-cloud interaction.}.
For example, a small amount of protons, which were
accelerated when the SNR was younger and the magnetic field was stronger, could
have escaped the acceleration region and are hitting the surrounding molecular clouds now
to produce dim hadronic gamma-ray emission, while most of the gamma rays are produced by either
electrons or protons which are currently being accelerated.
Assuming a Bohm-like diffusion
in magnetic turbulence with an average strength of $\sim0.1$~mG, 
the propagation length of PeV protons within $10^3$~yr
is estimated to be $\sim1$~pc, so that
a non-negligible fraction of such high-energy protons can exist 
inside and around the SNR to interact with dense targets.
Spatially-resolved spectroscopy with the better angular resolution of CTA 
will help us investigate such `multi-zone' structures.
Another possible scenario is the following. Both protons and electrons are
currently being accelerated, but protons are accelerated in regions with stronger
magnetic fields, while electrons are accelerated in regions with
weaker magnetic fields. In the strong magnetic field regions,
electrons are less efficiently accelerated, because the Alfven Mach number is
small due to the strong upstream magnetic field, 
resulting in a small injection rate of electrons \citep{hoshino02}.
Generally speaking, 
if a young SNR is producing CRs up to the knee energy, 
we naturally expect a dim hadronic component with gamma-ray energy more than
100~TeV unless the hadrons have escaped into the ISM during an earlier phase. 
Note that this statement is independent from the detailed acceleration
mechanism. Therefore, if such dim and hard hadronic component will be detected,
it will provide evidence for proton acceleration up to the knee energy.

\item{\it Time variation of maximum CR energies}\\
The CTA observatory will last for more than ten years. This motivates us to
measure any possible time variation of the gamma-ray spectrum over 
long time-scales. 
It has been shown that 
the maximum energy of CRs confined in SNRs depends on 
the SNR age during the Sedov phase \citep[e.g.][]{ptuskin03,ohira12b}.
Since the downstream magnetic field strength around the shock decays with the 
shock velocity, as the SNR ages it becomes harder to confine high energy CRs 
through pitch-angle scatterings near the shock. 
Accordingly, the maximum energy of CR protons should also decrease with time\footnote{
In the Sedov phase, the maximum energy of CR protons
should no longer be constrained by the 
acceleration time-scale but rather by 
the CR escape process.}.
%
The overall normalization of the gamma-ray spectrum may also 
change with time although its behavior is highly uncertain.
Assuming that the energy in cosmic-ray protons is proportional to
the integration of the kinetic energy flux of upstream matter 
passing through the shock front, 
\citet{dwarkadas13} derived a simple scaling of hadronic
gamma-ray luminosity which can be written as $L_\gamma\propto t^{5m-2m\beta-2}$,
where the SNR radius evolves as $R\propto t^m$ and the
surrounding medium has a power-law profile
in density $\rho\propto r^{-\beta}$.
Then, for a uniform surrounding medium ($\beta=0$),
we obtain $L_\gamma\propto t^{3}$ for $m=1$ (free expansion),
and $L_\gamma\propto t^{0}$ for $m=2/5$ (adiabatic expansion).
In the case of $\beta=2$ (an isotropic wind), 
we find $L_\gamma\propto t^{-1}$ for $m=1$ (free expansion),
and $L_\gamma\propto t^{-4/3}$ for $m=2/3$ (adiabatic expansion).
Another model predicts a secular flux increase at a few hundred GeV
at a level around 15\% over 10 years \citep{federici15}

In the following, we consider the simplest case in which only the 
cutoff energy of the gamma-ray spectrum changes with time.
The cutoff energy may vary $\sim10$~\% in 10--20~years.
To demonstrate this, we consider a power-law behavior for 
the time evolution of the escape-limited maximum energy, i.e., 
$E_{\rm max,p} \propto t^{-\alpha}$.
The value of $\alpha$ is important for understanding the average source spectrum of  
Galactic CRs which should be steeper than $E^{-2}$ \citep{ohira10}. 
Although there are theoretical studies on $\alpha$ 
\citep[e.g.,][]{ptuskin03,yan12},
its precise value has not yet been determined
observationally nor theoretically. 
To reproduce the spectrum of Galactic CRs, we can phenomenologically assume 
that $E_{\rm max,p} \approx 10^{15.5}$eV 
at the beginning of the Sedov phase ($t_{\rm Sedov}$), 
and that $E_{\rm max,p} \approx 1~{\rm GeV}$ at the end of the Sedov phase 
($\approx 10^{2.5}~t_{\rm Sedov}$), from which 
we can obtain $\alpha \approx 2.6$ \citep{ohira10}.
The age of RX~J1713.7$-$3946 is around $10^3~{\rm yr}$ (section 1.2). 
Then, the expected evolution rate of the maximum energy at this age
is estimated as
$\dot{E}_{\rm max,p}/ E_{\rm max,p} \approx -\alpha/t 
\approx -2.6 \times 10^{-3}~{\rm yr}^{-1}$.
This implies a decay of the maximum energy of about $5~\%$ over a period of $20~{\rm yrs}$.
If the observed gamma rays are produced by CR protons, 
the cutoff energy of the gamma rays is proportional to $E_{\rm max,p}$.
On the other hand, the leptonic gamma-ray scenario predicts the cutoff energy 
to be proportional to $E^2_{\rm max,e}$. 
Therefore, the time variation of the cutoff energy of gamma rays over $20$~yrs is expected to be about 
$5~\%$ and $10~\%$ for CR protons and electrons, respectively.
These estimations are based on the assumption of a smooth transition 
of $E_{\rm max,p}$ or $E_{\rm max,e}$ with a power-law scaling.
However,
RX~J1713.7$-$3946 is thought to be currently interacting with dense regions. 
If so, a collision between the SNR shock and density bumps causes
a sudden deceleration of the shock,
which results in a variability of $E_{\rm max,p}$ and/or 
$E_{\rm max,e}$ on shorter timescales.

For the CR electrons, if their maximum energy, $E_{\rm max,e}$, is also limited by their escape from the SNR, 
its time evolution should be the same as that of the CR protons. 
On the other hand, if it is limited by synchrotron cooling, 
its evolution should differ from that of the CR protons. 
A cooling-limited $E_{\rm max,e}$ near the shock can either decrease or increase with time 
depending on the evolution of the amplified magnetic field \citep{ohira12b}, 
for which shock-cloud interactions can play a significant role.

\end{itemize}


\section{Simulations}

In this section, we evaluate the feasibility of achieving the scientific objectives discussed above using a series of observation simulations for CTA.
The simulation software package {\it ctools} \footnote{version 00-07-01. See also http://ascl.net/1601.005 (ascl.1601.005)} \citep{jur13,jur16} 
that we use in this study is developed as an open source project aiming at a versatile
analysis tool for the broader gamma-ray astronomy community,
and is very similar to the {\it Science Tools}\footnote{http://fermi.gsfc.nasa.gov/ssc/data/analysis/software/} developed for the {\it Fermi}/LAT data analysis.
{\it ctobssim} is an observation simulator which generates
event files
containing the reconstructed incident photon direction in sky coordinates,
reconstructed energy, and arrival time
for each VHE photon detected in the field-of-view.
Simulated VHE sky images are produced by {\it ctskymap} using the simulated event files.
Event selection based on these parameters can be applied by {\it ctselect}
and event binning is performed by {\it ctbin}.
Unbinned or binned likelihood model fitting is executed through {\it ctlike}.

\subsection{Input data}

\subsubsection{Instrumental information and backgrounds}
 

To conduct CTA simulations and analyses, we used a set of instrument response functions (IRFs) equivalent to those for the public "2Q" array configuration available at the CTA website\footnote{https://portal.cta-observatory.org/Pages/CTA-Performance.aspx. (version 2015-05-05)}.
50-hr point source observations at a zenith angle of 20$^\circ$ are assumed to generate the IRFs.
We assumed here observations pointing towards the centre of the remnant.

We modeled the spatial distribution of the isotropic background,
which originates from gamma-like charged cosmic-ray events, 
as a two-dimensional Gaussian in the field-of-view 
(due to the radial dependence of acceptance) 
with a standard deviation $\sigma$ of 3$^\circ$ throughout these studies.
Since RX~J1713.7$-$3946 is located on the Galactic Plane, 
we also took into account the Galactic diffuse background (GDBG).
We conservatively assumed that the GDBG has a power-law spectrum without a cutoff
and estimated it by extrapolating the GDBD model {\it gll\_iem\_v05\_rev1.fit}\footnote{available at http://fermi.gsfc.nasa.gov/ssc/data/access/lat/BackgroundModels.html}
provided by the {\it Fermi}/LAT team \citep{gdbg16}.
Using only 6 data points above 0.1~TeV in the {\it Fermi}/LAT data up to $>513$~GeV, 
we derived a net GDBG flux within 1~degree around RX~J1713.7$-$3946 of $1.04\times 10^{-9}(E/0.1~{\rm TeV})^{-2.24}$~ph\,cm$^{-2}$s$^{-1}$TeV$^{-1}$, which is also roughly consistent with previous TeV observations \citep{hess14gd}.
With these assumptions, the GDBG exceeds the isotropic cosmic-ray background above a few TeV. 

\subsubsection{Gamma rays from RX~J1713.7$-$3946}
Here we assume that the gamma-ray spectrum is 
dominated by a leptonic component. 
As discussed in the previous section,
an additional hadronic component should exist.
By combining these two components, we made templates for our gamma-ray simulations as follows.

\begin{enumerate}
\item {\it Leptonic component}\\

We used the X-ray image of RX~J1713.7$-$3946 as a template that traces the leptonic gamma-ray
morphology. The X-ray image is extracted from the {\it XMM-Newton} archival data
consisting of multiple pointing observations to cover the entirety of this
remnant. The energy band of this image is restricted to 0.5--8.0~keV. 
The Non X-ray background is subtracted, and vignetting and exposure corrections
are applied. The pixel scale is 5$''$ and the image is smoothed with a
Gaussian function with a $\sigma$ of 3~pixels. Apparent point sources, likely
foreground stars or background extragalactic objects, as well as the central compact object (CCO) known as the neutron
star candidate 1WGA J1713.4$-$3949
\citep{lazendic03}, the brightest point source in this field, are excluded.

We consider a simplified case where the gamma-ray spectral shape is spatially invariant
over the two-dimensional image. 
The spectrum we adopt is based on the fitting results from H.E.S.S. data and can be written as
\begin{equation}
\frac{ {\rm d} N_e}{{\rm d}E }(E) = A_e \times \left(\frac{E}{\rm TeV}\right)^{-\Gamma _e}
\exp(-E/E_{\rm c}^{e})~~,
\end{equation}
where $A_e$ is a normalization constant, 
$\Gamma _e$ is the photon index ($2.04$) and $E_{\rm c}^{e}$ is the cutoff energy (17.9~TeV) \citep[see Table~4 of][]{hess07}.

\item {\it Hadronic component}\\
Based on CO and H{\sc i} observations, we obtain the total target gas column density (atomic $+$ molecular) \citep{fukui12}
and use it as a template that is assumed to trace the hadronic gamma-ray morphology.
The spatial boundary of the total target gas is defined using the apparent edge of the VHE gamma-ray emission shown in Aharonian et al. (2007). Additionally, we assumed that the spatial distribution of CR protons is homogeneous within the shell of the SNR.
Again, the gamma-ray spectral shape is assumed to be spatially invariant.
In this case the spectrum is written in a parametric form, i.e.,
\begin{equation}
\frac{ {\rm d} N_p}{{\rm d}E }(E) = A_p \times \left(\frac{E}{\rm TeV}\right)^{-\Gamma _p}
\exp(-E/E_{\rm c}^{p})~~,
\end{equation}
where $A_p$, 
$\Gamma _p$, and $E_{\rm c}^{p}$ are the normalization 
constant,
photon index, and cutoff energy, respectively.
\end{enumerate}
We adopt $\Gamma _p=2.0$ and $E_{\rm c}^{p}=300$~TeV as our fiducial parameters to picture a possible PeVatron accelerator.
In the following, we consider several cases with different values of $A_p/A_e$.
The absolute values of $A_e$ and $A_p$ are 
determined by requiring that the 
integration of the sum $N_e(E)+N_p(E)$ is equal to the total
photon flux between 1 and 10~TeV measured by H.E.S.S..

We generated gamma rays with energies between 0.3 and 100~TeV, which is narrower than the designed energy range of CTA in full-array configurations.
Since the spectral shape is concave below a few hundred of GeV \citep{abdo11rxj1713}, 
we restricted the simulation and analysis to above 0.3~TeV for simplicity. 
On the other hand, setting a high enough upper energy bound is extremely important with the aforementioned scientific goals in mind. 
Unfortunately, however, the IRF currently available supports simulations only up to $10^{2.2}\sim 158$~TeV, so that we simply cut our event generations at 100~TeV in this pilot simulation study.
In the near future the IRF will be updated and the energy coverage will be extended to 300~TeV and beyond. 
Working with such IRF and more detailed emission models are reserved for future work.

%
%
%
%
%
%
%
%
%
%
%

\subsection{Results and analysis}

\subsubsection{Gamma-ray image}
\begin{figure}[h]
\begin{center}
\includegraphics[width=168mm,clip]{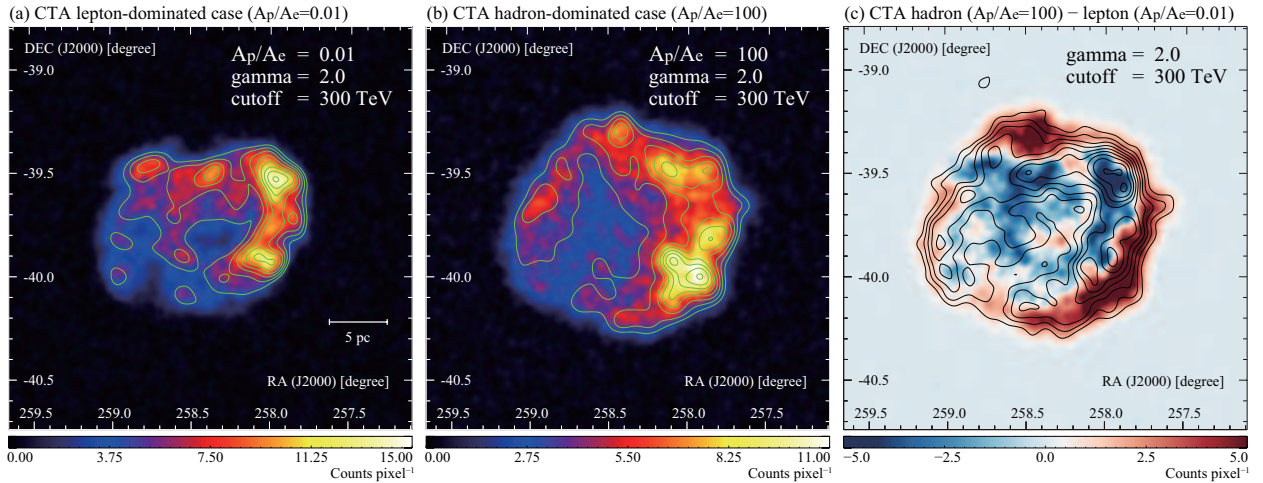}
\caption{
Simulated gamma-ray images of (a) $A_p/A_e = 0.01$ (lepton-dominated case) 
and (b) $A_p/A_e = 100$ (hadron-dominated case) with $\Gamma _p =2.0$ and 
$E_\mathrm{c}^{p}$ = 300 TeV. The green contours show (a) {\it XMM-Newton} X-ray intensity 
\citep[e.g., ][]{acero09} and (b) total interstellar proton column density 
\citep{fukui12} smoothed to match the PSF of CTA.
The subtracted image of (b)$-$(a) is also shown in (c).
The black contours correspond to the H.E.S.S. VHE gamma rays \citep{hess07}.
The unit of color axis is counts/pixel for all panels.
}
\label{fig1}
\end{center}
\end{figure}

We first look at the simulated images for CTA and study the possibility of using morphological information to determine the major component of the VHE gamma-ray emission.
Different gamma-ray images
in the energy range of 1--100~TeV are generated by changing $A_p/A_e$, 
the ratio between the hadronic and leptonic contributions.
Figures~\ref{fig1}a and \ref{fig1}b show the images 
for $A_p/A_e=0.01$ (lepton dominated; e.g., \citet{abdo11rxj1713}) and 100 (hadron dominated; e.g., \citet{fukui12}), respectively.  
Each image corresponds to 50~hr of observations with CTA.
As per our assumptions described above for the underlying templates, 
the lepton-dominated model (Figure \ref{fig1}a) shows a gamma-ray image 
that resembles the X-rays, and the hadron-dominated case (Figure \ref{fig1}b) 
delineates the ISM proton distribution including both CO and H{\sc i}.
The difference between them (Figure\,\ref{fig1}c) is significant 
as is evident from the subtraction between the two images.

To perform a quantitative evaluation, we employ the method of likelihood fitting (e.g., \citet{mattox96,cowan98,feigelson12}).
We focus here on the ability of CTA to determine the dominant emission component from the prospectively observed morphology.
We calculated the maximum log-likelihood, $L_e$ and $L_p$, by fitting the simulated images with the leptonic and hadronic spatial template, respectively.
We note that the templates used during the fits are the same as those used in the simulations,
meaning that a rather idealistic case is assumed.
To determine the log-likelihoods $L_e$ and $L_p$ individually,
we employ either the hadronic or leptonic template at a time in each fitting model
(regardless of the intrinsic $A_p/A_e$ ratio in the simulation data being fitted). 
We keep the normalization, photon index, and cutoff energy of the power-law as free parameters during the fitting process.
We also fit the data with the composite model containing both the leptonic and hadronic components 
in order to obtain the log-likelihood $L_{ep}$.
Then we calculated the differences $2(L_{ep} -L_e)$ and $2(L_{ep}-L_p)$ for various $A_p/A_e$, as summarized in Table\,\ref{table:morph}.
These differences in log-likelihood are large when the composite model is a significant improvement over the leptonic or hadronic models, 
and the difference is intended to be distributed approximately as a $\chi ^2$ variable with 3 degrees of freedom 
(for the extra parameters in the composite model). 
However this is only a rough indication as the conditions of Wilks' theorem are violated 
because the simple model is on the physical boundary of allowed parameters of the composite model \citep{wilks38}.
Table 1 indicates that the composite model is strongly favored over the simple models and that we can easily find the dominant component when the contribution of the second component is small. For example, when $A_p/A_e = 100$, $L_{ep}$ is very different from $L_e$ but nearly the same as $L_p$.
On the other hand, in the case where there is a more-or-less equal contribution from both particle populations (i.e., $A_p/A_e=1$), 
we found that the fitting tends to be biased toward the hadronic component. 
When $A_p/A_e=1$ in our model, the total number of hadronic photons is larger than that of the leptonic photons due to the differences in $E_c$, especially at higher energies.
Since the angular resolution becomes better at such higher energy range, 
the likelihood result expectedly favors a hadron-dominated scenario.
Nevertheless, we note that the case of $A_p/A_e > 1$ (hadron-dominated) shows a more spread-out structure in gamma rays than that of $A_p/A_e< 1$ (lepton-dominated). This trend is consistent with the latest H.E.S.S. results \citet{hess2016}. 
We conclude that CTA observations will be able to distinguish between hadronic and leptonic gamma rays based on morphological characterizations.

\begin{table}[h]
\begin{center}
\caption{ Differences in the maximum log-likelihood for spatial templates with various values of the $A_p/A_e$ ratio.}
\label{table:morph}

\begin{tabular}{lcc}

\hline
$A_p/A_e$\tablenotemark{\rm a} & $2(L_{ep}-L_e)$ \tablenotemark{\rm b}&   $2(L_{ep} - L_p)$\tablenotemark{\rm b} \\
\hline
0.01  & $16.8$ & $20085.8$\\
0.1   & $761.9$ & $15030.1$ \\
1.0   & $12280.2$ & $3494.1$ \\
10.0 & $27083.1$ & $171.9$  \\
100 & $31649.0$ & $0.9$ \\
\hline

\end{tabular}
\tablenotetext{\rm a}{$A_e$ and $A_p$ are normalization constants for the leptonic and hadronic component, respectively. The details are described in the text.}
\tablenotetext{\rm b}{$L_e$, $L_p$ and $L_{ep}$ are the maximum log-likelihoods for the leptonic, hadronic and composite templates, respectively}

\end{center}
\end{table}

\subsubsection{Spectrum}
In the case where the leptonic component is dominant, 
we can subsequently attempt to search for the ``hidden'' hard component with a hadronic origin as discussed above.
In order to evaluate the capability of CTA to achieve this interesting task, 
we further perform likelihood analyses over the wider energy band of $E>0.3$~TeV 
using 50~hr of simulation data with various assumed $A_p/A_e$ ratios.
The spatial templates and the spectral shapes in the fitting models are the same as those used originally for the simulation.
The fitting results are summarized in Table~\ref{tab:likelihood_various_ratio}.
We can significantly detect the hadronic component even for a small $A_p/A_e = 0.02$. 
However, the best-fit spectral shape for the hadronic component 
is generally slightly harder than the true input value of 2.0 for the cases with smaller ratios. 
This is probably because the dimmer and widely extended hadronic component is more easily confused with the background photons, especially at lower energies.
Indeed, when we simulated and analyzed without the GDBG nor the cosmic-ray background, 
the best-fit photon indices were converged consistent with the input value of $2.0$.
We thus refrain from drawing any strong conclusion on the detectability of the $A_p/A_e$ ratio at this point.

\begin{table}[h]
\begin{center}
\caption{ Resulting parameters of the likelihood analyses for the simulation data with various values of the $A_p/A_e$ ratio.
 \label{tab:likelihood_various_ratio}}
\begin{tabular}{lcccc}
\tableline \tableline
 $A_p/A_e$\tablenotemark{\rm a} &  $2 (L_{ep}-L_e)$\tablenotemark{\rm b} &  $\Gamma_{e}$ \tablenotemark{\rm c} & $E_c^e$ \tablenotemark{\rm d} & $\Gamma_p$\tablenotemark{\rm e} \\\tableline
 0.01 &    9.2 &    $2.03 \pm 0.01$ & $17.08 \pm 0.56$ & $1.73 \pm 0.26$ \\ 
 0.02 &   41.9 &    $2.03 \pm 0.01$ & $17.50 \pm 0.60$ & $1.86 \pm 0.12$ \\ 
 0.06 &  346.5 &    $2.04 \pm 0.01$ & $17.62 \pm 0.66$ & $1.90 \pm 0.04$ \\ 
 0.10 &  998.0 &    $2.03 \pm 0.01$ & $17.00 \pm 0.67$ & $1.93 \pm 0.03$ \\ 
\tableline
\end{tabular}
\tablenotetext{\rm a}{$A_e$ and $A_p$ are normalization constants for the leptonic and hadronic component, respectively. The details are described in the text.}
\tablenotetext{\rm b}{$L$ and $L_{\rm 0}$ are the maximum log-likelihoods for the model with/without the hadronic component, respectively.}
\tablenotetext{\rm c}{$\Gamma_{e}$ is a photon index for the gamma-ray spectrum of the leptonic component.}
\tablenotetext{\rm d}{$E_c^e$ is a cutoff energy for the gamma-ray spectrum of the leptonic component.}
\tablenotetext{\rm e}{$\Gamma_p$ is a photon index for the gamma-ray spectrum of the hadronic component.}
\end{center}
\end{table}

\begin{figure}[h]
\begin{center}
\includegraphics[width=0.6\linewidth]{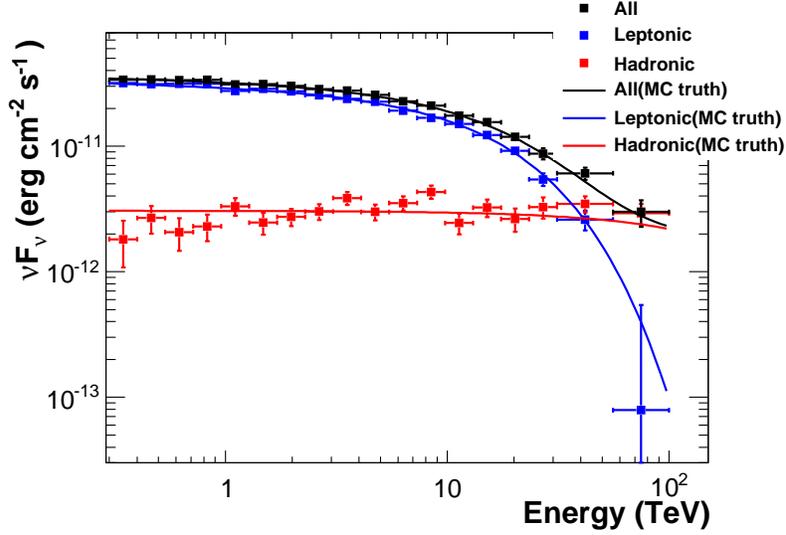}
\caption{
Spectral energy distribution of the gamma ray emission
obtained by analyzing the CTA simulation data for RX~J1713.7$-$3946.
The blue and red squares are the spectral points for the leptonic and 
hadronic spatial templates, respectively.
Vertical bars show statistical errors.
The black squares are the total fluxes of the leptonic and 
hadronic components.
The black vertical bars are the errors for the total fluxes 
obtained by adding the errors for two components in quadrature. 
The blue, red, and black solid lines show the input spectra for the leptonic component, 
the hadronic component, and the total, respectively.
\label{fig:bin_by_bin_spec}}
\end{center}
\end{figure}

\begin{figure}[h]
\begin{center}
\includegraphics[width=0.6\linewidth]{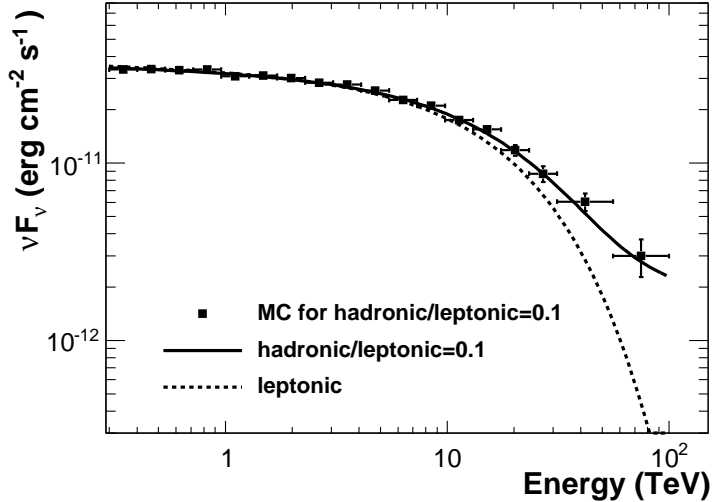}
\caption{
Comparison of the spectral energy distribution of the gamma-ray 
emission obtained by analyzing the CTA simulation data for RX~J1713.7$-$3946
with the models.
The black squares are the total of the fluxes for the leptonic and 
hadronic spatial templates.
Horizontal bars indicate the energy range the flux refers to.
Vertical bars show the errors obtained by adding the errors 
for two components in quadrature. 
The solid line shows the input spectra of gamma-ray simulation. 
The dotted line is for the model when the emission is purely leptonic.
\label{fig:comp_leptonic}}
\end{center}
\end{figure}

We proceed to perform maximum likelihood fits for the simulation data (with a ratio $A_p/A_e= 0.1$) for 18 logarithmically spaced energy bands 
spanning from 0.3~TeV to 100~TeV. Unlike our analysis above where we have to specify spectral shapes 
for the hadronic and leptonic components, in this way we can measure the spectrum independently of any 
spectral model assumption. 
Figure~\ref{fig:bin_by_bin_spec} shows the resulting spectrum from our ``bin-by-bin'' analysis
of the same 50~hr simulation data. 
For each spatial template, 
it is clear that our likelihood fits satisfactorily reproduce the simulated spectrum above 0.3~TeV. 
The data in each energy bin were fit by power-laws with a fixed index of 2.0.
As a final check, we compared the total gamma-ray flux points from the likelihood fit
over the 18 energy bins with our simulation model in which the emission is purely leptonic (see Figure~\ref{fig:comp_leptonic}).
Although it depends on the energy binning,
in the case of $A_p/A_e= 0.1$ and 50~hr of observation,
we can obtain the flux points above 30~TeV
deviating from the purely leptonic model
with a statistical significance level of $>$~3$\sigma$.
Thus, we reach the conclusion that 50~hrs CTA observation could detect at above $3\sigma$ the hard hadronic component with a ratio $A_p/A_e= 0.1$, under the idealistic conditions assumed here.

\subsubsection{Time variation of cutoff energy}
Detecting the time variation of the cutoff energy of the gamma-ray spectrum provides a clue to distinguish between different emission scenarios and CR acceleration theories.
We start off with the generation of simulated photons for a 100-hr observation of RX~J1713.7$-$3946.
The simulations are performed for three sets of intrinsic cutoff energies at 17.9, 19.7 and 16.1~TeV each.
The first set with $E_{\rm c} = 17.9$~TeV is tagged as our nominal case based on the current best-fit value suggested by H.E.S.S. data. 
The other two cases represent the possibility of a varying $E_{\rm c}$ in the coming years within the CTA mission lifetime. 
As a start, we consider a variation of $\Delta E_{\rm c}/E_{\rm c} = \pm 10\%$ to obtain an initial impression on how sensitive CTA will be to such fractional changes in the spectral cutoff.
Roughly $\sim 10$~yrs could be expected for $\pm 10$\% variation as discussed in Section 2.
Here we consider the purely leptonic scenario as an example (see Section 3.1 for details).

\begin{figure}[h]
\begin{center}
\includegraphics[width=0.7\linewidth]{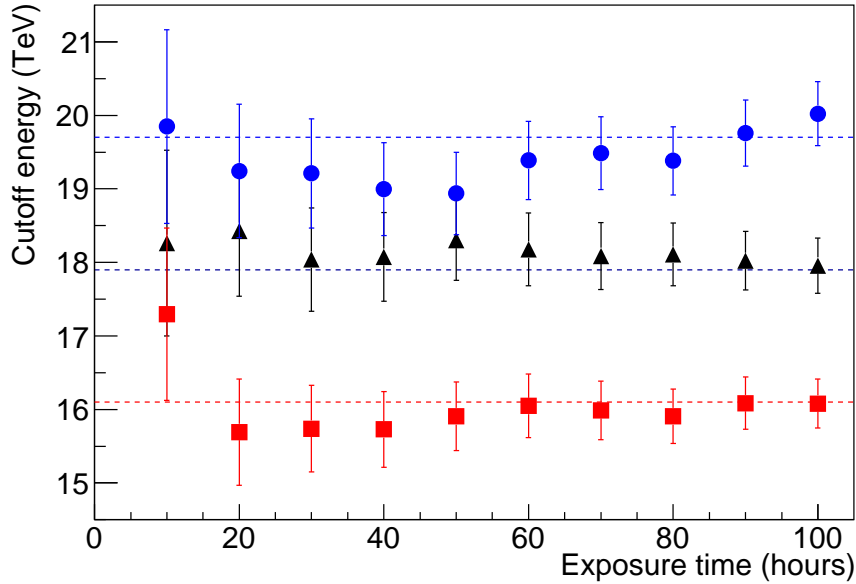}
\caption{Best-fit cutoff energies derived from our spectral analyses as a function of observation time. 
Circle, triangle and square markers correspond to the case of $\Delta E_{\rm c}/E{\rm c} = +10\%, 0\%$, and $-10$\%, respectively. 
Dashed lines indicate the true values of the simulations.}
\label{fig-Ec}
\end{center}
\end{figure}

For each case, we explore how the detectability depends on the total exposure time 
by extracting subsets from the full dataset with different exposures shorter than 100 hrs. 
This is achieved by applying a time-based cut using {\it ctselect}.
We then perform unbinned likelihood fitting for each of these subsets 
to derive the corresponding best-fit cutoff energies.
In the fitting, the normalization, photon index and $E^{e}_{\rm c}$ are all free parameters and are fitted simultaneously.
Figure\,\ref{fig-Ec} shows the best-fit $E^{e}_{\rm c}$ and their associated errors from our likelihood analyses.
As expected, the fitted $E^{e}_{\rm c}$ are closer to their true values with smaller uncertainties for longer exposures in all cases.

\begin{figure}[htp]
\begin{center}
\includegraphics[width=0.7\linewidth]{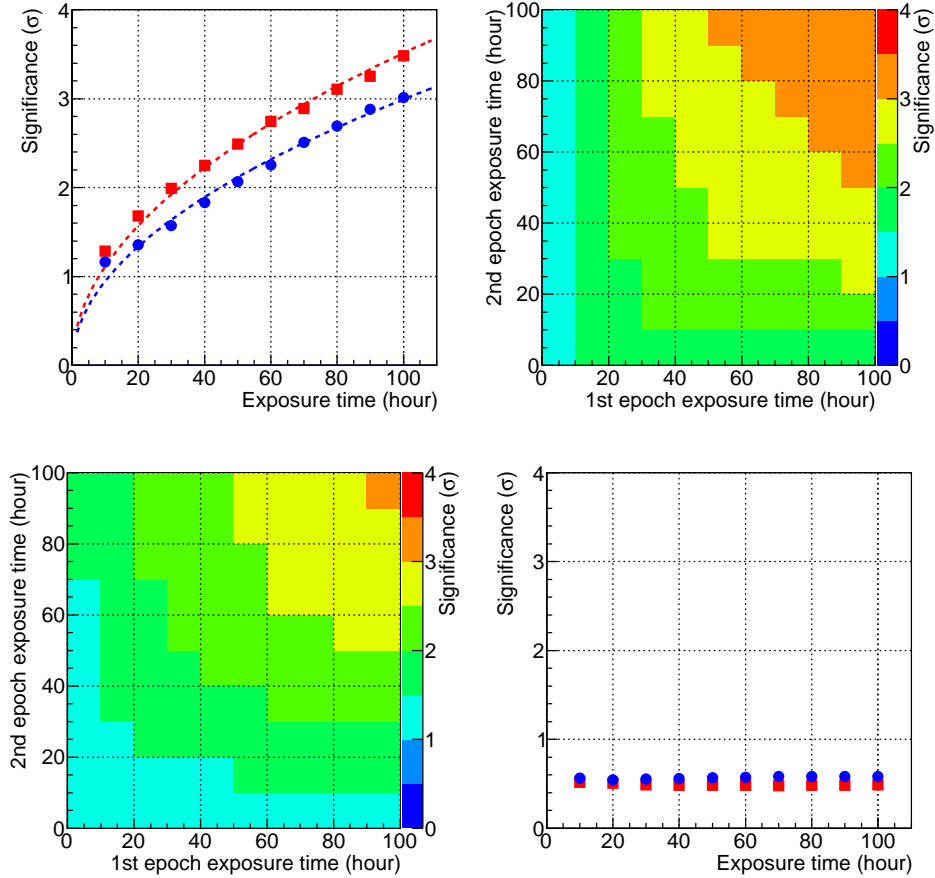}
\caption{Significance of the detected variation of $E_{\rm c}$ as a function of exposure time for three observation scenarios (see text for detail).
  Top left panel: The same exposure time for the two epochs of observations. Dashed lines represent the best-fit curve to each dataset. 
  Squares and circles represent results for the $\Delta E_{\rm c}/E_{\rm c} = -10$\% and $+10$\% case respectively.
  Top right panel: The results for various exposure times for the first and second dataset for the $\Delta E_{\rm c}/E_{\rm c} = -10$\% case.
  Bottom left panel: The same as the top right panel, but for the $\Delta E_{\rm c}/E_{\rm c} = +10$\% case.
  Bottom right panel: The scenario of using H.E.S.S. results for the first epoch \citep{hess07}.
  Squares and circles represent results for the $\Delta E_{\rm c}/E_{\rm c} = -10$\% and $+10$\% case respectively.
}
\label{fig-ecsigma}
\end{center}
\end{figure}

We then proceed to define a significance, $s$, for any observed $E^{e}_{\rm c}$ variation as
\begin{equation}
  s _{\pm} (t)= \frac{|E_\pm(t) - E_0(t)|}{\sqrt{\sigma _\pm^2(t) + \sigma _0^2(t)}}, 
\label{eq:spar}
\end{equation}
where $E_0$ and $E_\pm$ are the best fit $E_{\rm c}$ for the cases with the nominal value and $\pm 10\%$ variation, respectively. 
$\sigma _0$ and $\sigma _\pm$ are the corresponding errors.
To suppress statistical fluctuations, we repeat all of our simulations for 100 times and take the average of the calculated $s(t)$ from all runs.
In Figure\,\ref{fig-ecsigma} (Top left),  
our result indicates that a decrease in $E_{\rm c}$ with time is slightly easier to identify than an increase, which can be understood as follows. 
In the energy range of $E_{\rm c}$ ($\sim 10$~TeV), the sensitivity of CTA begins to drop with energy.
As a result, a lower cutoff energy can actually be measured more easily and precisely for a given exposure.
Our result is hence consistent with expectations.
We perform a simple fit to the points by a function $\propto \sqrt{t}$, where $t$ is the exposure.
If we observe $> 70$~hrs in the two epochs, we are able to achieve a $3\sigma$ detection for the $\Delta E_{\rm c}/E_{\rm c}=- 10\%$ case,
whereas $\sim 100$~hrs are necessary for the $+10\%$ case.

We can also estimate $s(t)$ from another point-of-view by considering the following observation scenario.
Suppose we will observe RX~J1713.7$-$3946 with different exposures during the first and the second epochs, 
how does the detection significance of a variation in $E_{\rm c}$ depend on the observation time in each epoch?
In this case, we can calculate 
\begin{equation}
s_\pm(t_1, t_2)= \frac{|E_\pm(t_2) - E_0(t_1)|}{\sqrt{\sigma _\pm^2(t_2) + \sigma _0^2(t_1)}},
\end{equation}
where $t_1$ and $t_2$ are the exposure time for the first and second epoch, respectively.
Our result is presented in Figure\,\ref{fig-ecsigma} (top right and bottom left panels).
Here we again arrive at the conclusion that an increase of $E_{\rm c}$ is harder to detect.
A longer exposure in the first year apparently makes it easier for us to detect the variation with a shorter observation in the next epoch.
Although several combinations of observation time are available for the 3$\sigma$ detection,
50-hr is the minimum required for the first observation period.

Since the full operation of CTA will begin roughly 10~years after the H.E.S.S. observations,
it is also of interest to know whether it is possible to detect any variation  
of $E_{\rm c}$ by comparing previous H.E.S.S. results and the upcoming CTA observations.
In this context, we can calculate $s(t)$ again by Equation\,(\ref{eq:spar}) fixing $E_0=17.9$~TeV and $\sigma _0=3.3$~TeV \citep{hess07}
and repeat the same procedure described above.
The result is plotted in Figure\,\ref{fig-ecsigma} (bottom right panel).
Owing to the large uncertainty of the H.E.S.S. results, however, 
the expected significance remains low and unmeaningful even with long CTA exposures up to 100~hrs.
This result consolidates the necessity of performing CTA observations of RX~J1713.7$-$3946
in at least two epochs separated by a $\sim$10-yr time interval.

\section{Summary and discussion}

In this paper, we have studied the feasibility and prospects for achieving a set of key scientific goals through CTA observations of RX~J1713.7$-$3946, based on simulations with exposures of 50-hrs or above .
We showed that a 50-hr observation is adequate to identify the dominant gamma-ray emission component, namely leptonic or hadronic, from the morphology of the SNR that is going to be revealed by CTA.
And we should be able not only to quantify both the leptonic and hadronic components but also to detect a possible hidden hard component at $>3\sigma$ level through spectral analyses if they are mixed with a ratio of $A_p/A_e \geq 0.1$.
Interestingly, we also found that CTA will be able to reveal fractional variations of the spectral cutoff energy over a timescale of $\sim 10$~yrs, for the very first time.
A variation of $\Delta E_{\rm c}/E_{\rm c} = \pm 10\% $ is found to be detectable
provided that an exposure time longer than 50~hr can be secured for the first epoch.
The result may indicate that the detection would be a tough task,
but at the same time the achievable science convinces us that it is worth the challenge.

The uncertainties of our results presented above are purely statistical.
The energy dispersion and uncertainty in the measured energy scale
can affect the fidelity of the spectral analyses especially in the measurement of $E_{\rm c}$.
We confirmed that the significance of the $E_{\rm c}$ variations due to the predicted energy dispersion would be negligible,
while the expected energy scale uncertainty would cause $\sim \pm 1\sigma$ deviations, respectively.
We also confirmed that 
the contamination of the very faint thermal X-ray to the leptonic template 
and variations of the hadronic template due to the uncertainty of the radio observations
are both negligible to the results. 
\citet{katsuda15} reported the detection of very faint thermal components 
but their spectral analysis region is limited only in the very center of this SNR with the radius of $5.3^\prime$. 
They also presented the softness map and the analysis is performed on one of the softest region 
where the thermal component is brighter than other area. 
Even such region, the flux of the thermal emission is far less than 10\% of the synchrotron. 
Concerning the radio map, the uncertainty of the intensity is estimated as less than 10\%. 
When we modified the hadronic image template by multiplying a random factor conservatively between $\pm 10$\% to every image pixel, 
the results are very similar and differences are typically less than $1\sigma$.
%
%
%
%
%
%
%
Estimating other systematic uncertainties for our analysis is not trivial and can be highly model-dependent. 
Possible effects due to a deviation from the actual PSF profile with a tail\footnote{as presented and discussed in \citet{aleksic16} for the case of MAGIC telescopes.} 
may exist, since the PSF was assumed simply as a Gaussian in our simulations.
Since RX~J1713.7$-$3946 is located at the Galactic Plane, 
the source is expected to be contaminated by the Galactic diffuse gamma-ray emission 
which constitutes the majority of the background events (i.e., ``noise'' for our purpose).
Possible existence of currently unknown TeV sources nearby, such as the CCO 1WGA J1713.4$-$3939 for instance, can contribute to the systematic errors
in such a crowded region on the Galactic Plane.



On the other hand, CTA will measure gamma rays with energies far below 0.3~TeV,
where the {\it Fermi}/LAT and H.E.S.S. data points connect.
If the gamma rays are mostly of a leptonic origin and the 
magnetic field strength $B$ is around 10 or a few tens of $\mu$G,
which has been inferred from theoretical studies 
\citep[e.g.,][]{yamazaki09,ellison10,lee2012},
we expect that the electron spectrum possesses a cooling break \citep{longair94} at
$\approx7.5$~TeV $(B/40~\mu{\rm G})^{-2}(t_{\rm age}/10^3~{\rm yr})^{-1}$,
where $t_{\rm age}$ is the age of RX~J1713.7$-$3946.
In this case the gamma-ray spectrum must also exhibit a corresponding break at 
$E_b\approx0.3$~TeV $(B/40~\mu{\rm G})^{-4}(t_{\rm age}/10^3{\rm yr})^{-2}$ (since $E_\gamma \propto E_e^2$ for inverse-Compton emission in the Thompson regime), if the particle injection is impulsive.
Future identification of such a cooling break feature in the
gamma-ray spectrum by CTA will equip us with a powerful and independent tool to measure the value of $B$, 
which would bring important constraints on the particle acceleration mechanism 
at high Mach-number collisionless shocks.


Our present study is based on a fairly simplified input model, but 
can be extended to include more physics motivated by theoretical models.
Some possibilities are listed in the following, and their incorporation in our model is reserved for future works.

(1) {\it Hadronic model based on MHD simulations}:
The shock interaction with dense clouds excites turbulence, which amplifies the 
magnetic field up to the mG order \citep{inoue12,sano13,sano14}. 
These results may be
 crucial for our better understanding of the gamma ray and 
X-ray images of SNRs. 
Under these circumstances, the CR electrons will be significantly 
affected via synchrotron cooling for magnetic fields greater than 
100~$\mu$G, and the gamma ray image will follow if the 
leptonic gamma-ray production is important. 
In order to fully understand such effects on the gamma-ray images, 
we need to incorporate the shock-cloud interaction by utilizing 
the ISM distribution with a reasonably high angular resolution as well as  
detailed numerical simulations.
Using the full performance of CTA by lowering the energy threshold could be essential in these studies,
since more apparent differences between the leptonic and hadronic scenarios are expected.
(2) {\it Highest energy CRs from SNRs}:
Accelerated particles eventually escape from the SNR forward shock to  
become Galactic CRs. According to recent studies, 
the highest energy CRs start to escape from the SNR 
at the beginning of the Sedov phase \citep{ptuskin05, ohira10, ohira12}. 
The escaped CRs make a gamma-ray halo, 
which is more extended than the SNR shell \citep{gabici09, fujita10, ohira11}. 
Spectral measurements of gamma-ray halos will provide information 
on the highest energy CRs accelerated in the SNR in the past.
Observed profiles of the gamma-ray halo will also constrain the diffusion coefficient of CRs and density structure in the ambient medium \citep{fujita11, malkov13}.
(3) {\it Hydrodynamical models and spectral modifications via nonlinear effects}:
More sophisticated models taking into account nonlinear DSA processes
predict realistic CR proton electron spectra
\citep[e.g.][]{zirakashvili10,lee2012}, both trapped and escaped, 
with a full treatment of their time evolution and 
spatial distribution in a hydrodynamic calculation for RX~J1713.7$-$3946. 
The comparison of these models with CTA results will provide important 
new insights on the very complex nonlinear DSA mechanism at young SNRs.     

\acknowledgments

This research made use of {\it ctools}, a community-developed analysis package for Imaging Air Cherenkov Telescope data. 
{\it ctools} is based on GammaLib, a community-developed toolbox for the high-level analysis of astronomical gamma-ray data.

We gratefully acknowledge financial support from
the following agencies and organizations:

Ministerio de Ciencia, Tecnolog\'ia e Innovaci\'on Productiva (MinCyT),
Comisi\'on Nacional de Energ\'ia At\'omica (CNEA), Consejo Nacional
de Investigaciones Cient\'ificas y T\'ecnicas (CONICET), Argentina;
State Committee of Science of Armenia, Armenia;
Conselho Nacional de Desenvolvimento Cient\'{i}fico e Tecnol\'{o}gico (CNPq),
Funda\c{c}\~{a}o de Amparo \`{a} Pesquisa do Estado do Rio de Janeiro (FAPERJ),
Funda\c{c}\~{a}o de Amparo \`{a} Pesquisa do Estado de S\~{a}o Paulo (FAPESP), Brasil;
Croatian Science Foundation, Croatia; 
Ministry of Education, Youth and Sports, MEYS LE13012, LG14019, 7AMB14AR005 and the
Czech Science Foundation grant 14-17501S, Czech Republic;
Ministry of Higher Education and Research, CNRS-INSU and CNRS-IN2P3, CEA-Irfu, ANR,
Regional Council Ile de France, Labex ENIGMASS, OSUG2020 and OCEVU, France;
Max Planck Society, BMBF, DESY, Helmholtz Association, Germany;
Department of Atomic Energy, Department of Science and Technology, India;
Istituto Nazionale di Astrofisica (INAF),
Istituto Nazionale di Fisica Nucleare (INFN), MIUR, Italy;
ICRR, University of Tokyo, JSPS, Japan;
Netherlands Research School for Astronomy (NOVA),
Netherlands Organization for Scientific Research (NWO), Netherlands;
The Bergen Research Foundation, Norway;
Ministry of Science and Higher Education, the National Centre for Research and 
Development and the National Science Centre, Poland;
Slovenian Research Agency, Slovenia;
MINECO support through the National R+D+I, Subprograma Severo Ochoa, CDTI 
funding plans and the CPAN and MultiDark Consolider-Ingenio 2010 programme, Spain;
Swedish Research Council, Royal Swedish Academy of Sciences, Sweden;
Swiss National Science Foundation (SNSF), Ernest Boninchi Foundation, Switzerland;
Durham University, Leverhulme Trust, Liverpool University, University of Leicester,
University of Oxford, Royal Society, Science and Technologies Facilities Council, UK;
U.S. National Science Foundation, U.S. Department of Energy,
Argonne National Laboratory, Barnard College, University of California,
University of Chicago, Columbia University, Georgia Institute of Technology,
Institute for Nuclear and Particle Astrophysics (INPAC-MRPI program),
Iowa State University, Washington University McDonnell Center for the Space Sciences, USA.

The research leading to these results has received funding from the
European Union's Seventh Framework Programme (FP7/2007-2013) under grant
agreement n$^\mathrm{o}$ 262053.

 We thank S. Katsuda for providing the original \textit{XMM-Newton} image.


\end{document}